\documentclass[
prl
,showpacs
,showkeys
,aps
,amsfonts
,amssymb
,byrevtex
,pdftex  
,twocolumn,nofootinbib
]{revtex4-1}

\usepackage{graphicx}  
\usepackage[export]{adjustbox}
\usepackage[titletoc]{appendix}
\usepackage{amsmath, amssymb, bm}   
\usepackage[dvips]{color}
\usepackage{natbib}
\usepackage{hyperref}
\usepackage{float}
\usepackage{array}
\newcolumntype{L}[1]{>{\raggedright\let\newline\\\arraybackslash\hspace{0pt}}m{#1}}
\newcolumntype{C}[1]{>{\centering\let\newline\\\arraybackslash\hspace{0pt}}m{#1}}
\newcolumntype{R}[1]{>{\raggedleft\let\newline\\\arraybackslash\hspace{0pt}}m{#1}}

\def\){\right)} 
\def\({\left(} 
\def\]{\right]} 
\def\[{\left[}
\def\gm{{$g$-mode}}
\def\bv{{Brunt-V\"ais\"al\"a}}

\begin{document}

\title{
 Lifting the veil on quark matter in compact stars with core g-mode oscillations}

\author{%
Wei Wei
}
\email{weiwei1981@mail.hzau.edu.cn}

\affiliation{College of Science, Huazhong Agricultural University, Wuhan, Hubei, P.R.China}

\author{%
Megan Barry, Thomas Kl\"ahn, Prashanth Jaikumar
}

\email{prashanth.jaikumar@csulb.edu}
\affiliation{Department of Physics $\&$ Astronomy,
California State University Long Beach, Long Beach, CA
  90840, U.S.A. }


\begin{abstract}

Compact stars containing quark matter may masquerade as neutron stars in the range of measured mass and radius, making it difficult to draw firm conclusions on the phase of matter inside the star. The sensitivity of core \gm\, oscillations to the presence of a mixed phase may alleviate this difficulty. In hybrid stars that admit quark matter in a mixed phase, the \gm\, frequency rises sharply due to a marked decrease in the equilibrium sound speed. Resonant excitation of \gm s can leave an imprint in the waveform of coalescing binary compact stars. We present some analytic and numeric results to assess the sensitivity displayed by \gm\, oscillations to quark matter in a homogeneous or mixed phase and find that the \gm\, can probe the fraction of quark matter inside the star.

\end{abstract}


\maketitle
\section{I. Introduction} The discovery by the Advanced LIGO and Advanced VIRGO collaborations of the binary neutron star merger GW170817~\cite{GW170817} opens a new observational window into compact star properties. Many recent works~\cite{Radice,Poles,AbbottEOS,De,Chatz, Malik,Tews:2018chv,Zhu,Christian:2018jyd,Li:2018ayl} have explored constraints on the neutron star equation of state (EoS) using tidal polarizabilities extracted from gravitational waveforms during the late inspiral phase. It appears possible, though not conclusive, that one or both of the component stars in the merger could be hybrid stars; that is, they support a phase transition to quark matter at high density~\cite{Paschalidis,Nandi}. In effect, the so-called ``masquerade" problem~\cite{Alford04} for compact stars persists:  a hybrid star with quark matter in its interior is indistinguishable from an ordinary neutron star based on current observational status, especially if quark matter is in a mixed phase with hadronic matter. A confirmation of this possibility was made in~\cite{Wei1} for nuclear to 2-flavor, 3-flavor and sequential flavor transitions. It has been suggested~\cite{Brillante,Flores13,Sotani10,Sotani12} that mapping out the radial or non-radial oscillation mode frequencies can provide a clear distinction between neutron and hybrid stars, with only a weak dependence on the poorly known EoS of the quark phase. Given this idea, and that non-radial oscillations couple to gravitational waves, we examine a diagnostic of the phase structure of matter inside neutron stars: the core \gm\, oscillations~\cite{RG,Miniutti02}. We find that these modes are sensitive to the presence and the proportion of quark matter inside neutron stars, similar to the conclusions in~\cite{Dommes15,Yu17}, which focused on hyperons. However, the appearance of hyperons does not involve any phase transition, so their effect on the \gm\, is less dramatic. If core \gm\, oscillations resonantly excited by tidal forces during the inspiral~\cite{Lai94} are detected, either directly with third generation detectors~\cite{Yu17} or indirectly through gravitational wave phasing~\cite{Andersson18}, that could finally lift the veil on quark matter inside neutron stars and resolve the masquerade problem.

In this paper, our focus is on the \gm\, characteristics in the quark-hadron mixed phase through a theoretical analysis of the restoring force (buoyancy) and damping. Typically, buoyancy arising from thermal and/or chemical stratification inside the star (core or crust) drives the \gm\,. It can also arise due to sharp density discontinuities in the star, as for first-order phase transitions~\cite{Miniutti02,Kruger15} without finite size effects. Core \gm\, oscillations arising from chemical stratification carry the imprint of the fluid's composition, a feature that can potentially be exploited by gravitational wave detectors operating in the (0.1-1) kHz regime~\cite{Yu17}. The \gm\, is driven by the mismatch between mechanical and chemical equilibrium rates of a displaced fluid parcel, expressed by the difference between local equilibrium and adiabatic sound speed respectively, i.e., the \bv\, frequency. Compared to a pure phase, a mixed phase system (ignoring surface and Coulomb forces) is more compressible due to its ability to distribute conserved charges globally. The drop in the equilibrium sound speed upon onset of the mixed phase is reflected in an increase of the \gm\, frequency. This is the basic result we exploit in this paper to characterize the \gm\, as a diagnostic for the phase transition to quark matter. 

 The \gm\, for $(n,p,e^-)$ matter with or without additional leptonic and hadronic species has been addressed in previous works~\cite{RG,Lai94,Kantor14,Yu17,Dommes15,Zhou18,Pass16}. For quark matter, model-dependent numerical studies have been reported~\cite{Sotani10,Fu17,FL,Sandoval18}, but none are in the context of a mixed phase (continuous phase transition). We take a simpler but more general approach that allows for analytic estimates of the \bv\, frequency in quark matter, and reveals the sensitivity of the \gm\, to the onset and the proportion of the quark phase. The \gm\, frequency vanishes in non-interacting and massless two and three (or any $N_f$) flavor quark matter, but can appear in any of the following realistic situations: non-zero quark mass, inclusion of interactions, a
 quark-hadron mixed phase. We illustrate these three cases separately for the sake of simplicity. The first and second are treated with analytic approximations, whereas for the third, we employ a common parameterization, where hadronic matter is described by a member of the family of PAL EsoS~\cite{PAL}, and quark matter is described by the vBag EoS (vector-enhanced Bag model)~\cite{Klahn1}. Within acceptable parameter ranges of these models, we find a steep rise in the \gm\,frequency upon the appearance of a mixed phase. We briefly discuss how this can impact tidal resonance phenomena in binary neutron star mergers where one or both components are hybrid stars with a (mixed-phase) quark matter core, and whether the effect can survive mode damping. A more detailed study of \gm\, resonant coupling to dynamical tides in neutron stars and the subsequent impact on gravitational wave phasing during inspiral is left to future work.
 
 This paper is organized as follows: sec II describes how core \gm\, oscillations probe the phase structure of compact star interiors, sec III contains analytic results for the \bv\, frequency in models of interacting quark matter, sec IV gathers numeric results displaying the \gm\, jump at the onset of the mixed phase, sec V presents estimates of \gm\, damping times, followed by conclusions and discussion in sec VI.
\section{II. $g$-mode oscillations}

The $g$-modes arising from chemical stratification are quite sensitive to the composition of dense matter. Therefore, they may be a better probe of the EoS than the $f$ and $p$-modes. For example, the \gm\, frequency depends on the proton fraction which is affected by the nuclear symmetry energy. The latter determines important physical quantities such as the compact star's radius, its tidal deformability and neutrino emission thresholds~\cite{Sahoo2016,Zhang2018,Krastev18,Haensel91,Gandolfi11}. The symmetry energy also plays a key role in the properties of terrestrial nuclei, such as neutron skin thickness and dipole polarizabilities~\cite{Cao15,Dong15}. As such, theoretical studies of the $g$-modes add to the list of diagnostics of dense matter properties coming from other phenomena in nuclear astrophysics. We emphasize that the \gm\, addressed in this work is different from the crustal~\cite{Finn87}, thermal~\cite{Stroh93} and discontinuity~\cite{McD90} \gm s, since we assume a continuous phase transition without a density discontinuity.

In order to determine the \gm\,spectrum, we first construct the stellar structure using General Relativity (TOV equations). To simplify the linearized fluid perturbation equations from which we calculate the frequency of the \gm\,, we employ the Newtonian and Cowling approximations, neglecting the back reaction of the Newtonian gravitational potential. While this is not strictly consistent with the fully relativistic treatment of the background structure, the impact of these simplifying approximations is not severe, typically only affecting the frequencies of the $p$-mode and \gm\, at the 5-10\% level~\cite{Grig}. However, the $f$-mode frequencies at low angular quantum number can be more sensitive to the the Cowling approximation. To go beyond the Cowling approximation involves a considerable complication since the fluid equations must be treated in full General Relativity.  While this is essential for a self-consistent calculation of gravitational waves, here we are only trying to obtain the approximate trend in the frequency as a function of stellar parameters, not the explicit wave forms, for which the Cowling approximation is sufficient. 

Accordingly, the system of equations used to compute \gm\, frequencies in  the neutron star are given by~\cite{Fu17,LB}

\begin{eqnarray}
\label{eq:1}
    \frac{\partial}{\partial r}(r^2\xi_r) &=&\left[\frac{l(l+1)}{\omega^2}-\frac{r^2}{c_s^2}\right]\left(\frac{\delta\rho}{\rho}\right)\\
\frac{\partial}{\partial r}\left(\frac{\delta\rho}{\rho}\right) &=& \frac{\omega^2-\omega_{BV}^2}{r^2}(r^2\xi_r)+\frac{\omega_{BV}^2}{g}\left(\frac{\delta\rho}{\rho}\right)
\label{eq:2}
\end{eqnarray}

where $\xi_r$ is the radial component of the fluid perturbation, $\delta\rho$ the Eulerian density perturbation, and the \bv\, frequency 

\begin{equation}
\label{BV}
    \omega_{BV}^2 \equiv N^2 = g^2\left(\frac{1}{c_e^2}-\frac{1}{c_s^2}\right) 
\end{equation}
 depends on the equilibrium ($c_e$) and adiabatic ($c_s$) sound speeds. The solution of the system of eqs.(\ref{eq:1}) \& (\ref{eq:2}) under relevant boundary conditions, viz., regularity at the stellar center ($r$$\rightarrow$0) and vanishing of the Lagrangian pressure variation $\Delta p = c_s^2 \Delta \rho$ at the surface, can exist only for discrete values of the mode frequency $\omega$. While we present numerical results for the $f$- and $p$-modes as well, our theoretical focus in this work is on the $l$=2 $g$-modes. 
 
 Eqs.(\ref{eq:1}) and (\ref{eq:2}) can be solved numerically once the background stellar configuration is specified. However, one can perform a local analysis of these equations in the eikonal approximation, which yields~\cite{RG}
 
\begin{equation}
    \omega^2 \approx \frac{l(l+1)}{(kr)^2+l(l+1)} N^2
\end{equation}
 
Convectively stable $g$-modes exist for $N^2 >0$, implying that

\begin{equation}
\label{csce}
    c_s^2-c_e^2=-\left(\frac{\partial P}{\partial x}\right)_\rho\left(\frac{dx}{d\rho}\right) > 0
\end{equation}
 
should be fulfilled, where $x$ is the proton fraction, which equals the electron fraction $Y_e = n_e/n_B$ in charge neutral $(n,p,e^-)$ matter. 

\section{III. Analytic Estimates for Simple Models of Dense Matter}

The sound speeds and \gm\, frequency can be approximately calculated analytically in a few simple, and surprisingly, even interacting models of nuclear or quark matter. As an example, we compute these quantities in interacting two flavor quark matter based on the vBag model~\cite{Klahn1}, which has been recently introduced to reconcile the missing features of the perturbative or thermodynamic Bag model (no chiral symmetry breaking) and NJL-type models (no confinement) within a single non-perturbative picture. This model is similar in spirit, but different in details than the more recent vMIT model~\cite{Gomes18}. The vBag model has proved to be versatile, with astrophysical applications such as mixed phases in neutron stars, protoneutron stars and supernova explosions, as demonstrated in recent works~\cite{Klahn2,Fischer,2017arXiv171208788F}. The purpose of studying these simple models is to emphasize the key quantities that determine the occurrence of stable $g$-modes, typically the nuclear or quark symmetry energy. 

\begin{figure}[H]
\includegraphics[height=60mm, width=90mm,left]{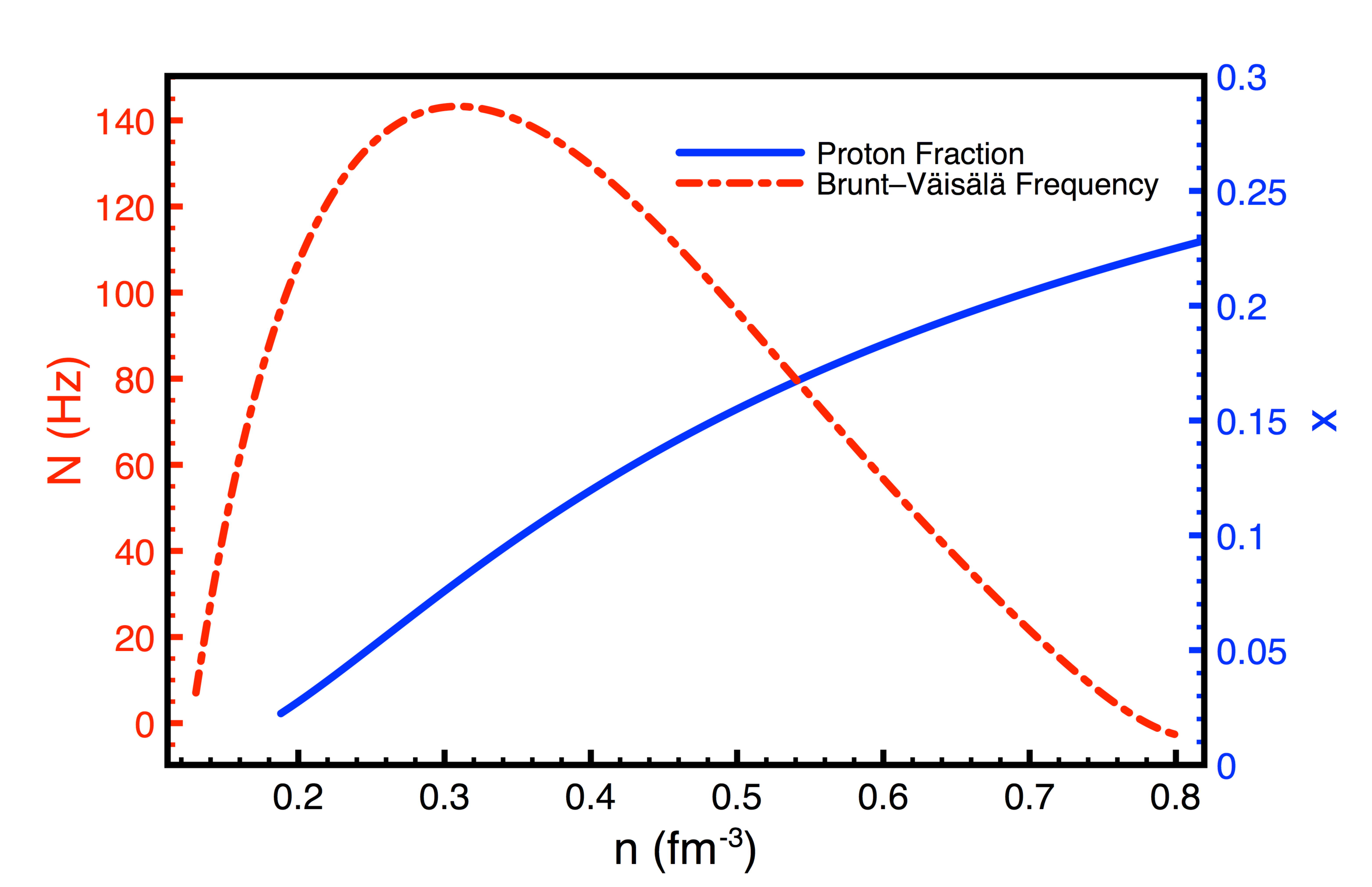}\\
\caption{\bv\, frequency $N$ (Hz) and proton fraction $x$ for the SL23 nuclear EoS across a typical density range from the core to inner crust boundary of the neutron star.}
\label{sl23fig}
\end{figure}

Before we list the analytic results, we emphasize the importance of the symmetry energy in nuclear/quark matter to the $g$-mode. Employing the widely used functional form for the nuclear contribution to the energy per baryon~\cite{PAL,WFF}, $E_b(n,x) \approx E_0(n) + E_s(n)(1-2x)^2$, it follows that

\begin{equation}
\frac{x}{(1-2x)^3}=\frac{64E_s(n)^3}{3\pi^2n}
\end{equation}

for $(n,p,e^-)$ matter in $\beta$ equilibrium, where $E_s(n)$ is the symmetry energy of uniform matter, and the proton fraction $x$=$x_e$ by charge neutrality. The extension to matter with muons is straightforward.
Following the techniques in~\cite{Lai94}, one can show that, in this parabolic approximation

\begin{equation}
N\approx 2\left(\frac{g}{c_e}\right)\left(\dfrac{x}{3}\right)^{1/2}\frac{(3nE_s^{'}-E_s)}{\sqrt{E_s(\frac{10}{9}E_0+2nE_s^{'}+n^2E_S^{''})}}
\end{equation}
 
where primes denotes density derivatives. It is clear from this expression that the \gm\, spectrum warrants further investigation to determine its sensitivity to properties of the nuclear medium. For the non-interacting $(n,p,e^-)$ gas, $N\approx (g/c_e)(3x/7)^{1/2}$, which is consistent with other approximate estimates in the literature~\cite{Lai94,RG}. Here, we have used natural or Planck units $\hbar$ = $c$ = $1$. The \bv\, frequency in the non-interacting case $N\sim$ 100 Hz. Extensions to various parameterized models of the nuclear interaction have been considered in~\cite{Lai94,RG,Fu17}. In our numerical calculations, we use the nuclear EoS SL23~\cite{PAL,Wei1}. The proton fraction and \bv\, frequency as a function of baryon density is displayed for EoS SL23 in Fig.\ref{sl23fig}. 

Since we are interested in the impact of quark matter on the \gm\,, we consider some simple models of quark matter where the \gm\, frequency can be estimated analytically. For non-interacting two flavor quark matter, charge neutrality and $\beta$-equilibrium provide two constraints on the up and down quark chemical potentials $\mu_u$ and $\mu_d$. This renders the electron fraction $x_e$ independent of the baryon density, as $x_e$ acquires the fixed value $\approx 0.0056$, implying that $\omega_{BV}$=0, and the absence of \gm\, oscillations. This situation is not realistic however, and we can imagine three different ways in which composition gradients can appear in quark matter. Firstly, the moderately heavy strange quark can appear at high density, modifying the charge neutrality condition such that $\mu_e$$\approx m_s^2/(4\mu_q)$, where $\mu_q$ is the quark chemical potential. Consequently, the electron fraction depends on density, providing the necessary variation of composition. Secondly, quark-quark interactions can generate $N$$\neq$0, which we demonstrate in the vBag model in the two-flavor sector. In this particular model, repulsive vector interactions that provide for hybrid stars as heavy as 2$M_{\odot}$ determine the electron fraction. Thirdly, even though non-interacting homogeneous two-flavor quark matter has a fixed electron fraction, $N$$\neq$0 can occur when such matter is part of a mixed phase with nuclear matter, where the pressure varies smoothly as the quark fraction grows. We model this case using Wood's relation~\cite{bubble} for a mixture of compressible fluids and show that it can support $g$-modes. In a more detailed numerical analysis in section IV, we will employ the Gibbs construction for the mixed phase and compute the \gm\, frequency therein.  

 For non-interacting massless two-flavor quark matter in the corresponding parabolic approximation for the energy per baryon of the quarks $E_q(n,x_e)$, the isospin asymmetry is given by $\delta=1-2x_e=3(n_d-n_u)/(n_d+n_u)$. The quark symmetry energy $E_s(n) \propto n^{1/3}$ which leads to a fixed value for $x_e\approx 0.0053$, within 5\% of the value obtained previously with no approximation to $E_q(n,x_e)$. In quark matter with massive strange quarks, the symmetry energy depends additionally on the fraction of strange quarks $x_s$=$n_s/n$~\cite{ChuChen},  wherefore $E_q(n,\delta,x_s) \approx E_0(n,x_s) 
+ E_s(n,x_s)\delta^2$ leads to a system of two equations that determine $x_e(n)$ and $x_s(n)$.

\begin{eqnarray}
\label{sym3flavor}
\frac{x_e}{(1-2x_e-2x_s/3)^3}&=&\frac{64E_s(n)^3}{3\pi^2n}\\
2-x_e-x_s&=&\left(x_s^2+\frac{m_s^2}{(\pi^2n)^{2/3}}\right)^{3/2}
\label{xe3flavor}
\end{eqnarray}

Therefore, the $g$-mode frequency, which directly involves the gradient of the composition with baryon number (or energy) density via eq.(\ref{csce}), serves as a probe of the symmetry energy.

Let us now address the three examples that provide for $g$-modes in quark matter. The first of them involves introducing the strange quark at sufficiently high density. Within the thermodynamic Bag model (henceforth, tdBag) with a Bag constant $B$, we can write the EoS for quark matter $P(\rho)$ to leading order in the strange quark's current mass $m_s$ as

\begin{equation}
P(\rho)=\frac{1}{3}(\rho-4B) -\frac{m_s^2}{3\pi}\sqrt{\rho-B}   
\end{equation}

where $P$ is the pressure and $\rho$ the energy density. We can ignore the tiny contribution of the electrons to the pressure, of order $m_s^8/\mu_q^4$. Effectively, this means the adiabatic sound speed is given by 

\begin{equation}
c_s^2=\frac{1}{3}-\frac{m_s^2}{6\pi\sqrt{\rho-B}}  
\end{equation}

Eq.(\ref{csce}) can be recast as~\cite{Lai94}

\begin{equation}
\label{csce2}
    c_s^2-c_e^2=-\left(\frac{n}{P+\rho}\right)\left(\frac{\partial P}{\partial x_e}\right)\left(\frac{dx_e}{dn}\right) 
\end{equation}
For the non-interacting case, an approximate solution to eqs.(\ref{sym3flavor}) and (\ref{xe3flavor}) that is accurate to better than a few percent for densities of interest yields $\mu_e\approx m_s^2/4\mu_q$. This implies that $x_e$ = $n_e/n\propto m_s^6/n^2$ so that $(dx_e/dn)=-2(x_e/n)$. Since $\mu_e/\mu_q \ll 1$, we can also approximate the quark pressure for homogeneous three-flavor quark matter as 

\begin{equation}
P(\mu_q,\mu_e)=P_0(\mu_q)-n_Q(\mu_q)\mu_e+{\cal O}(\mu_e^2)    
\end{equation}

with the charge density $n_Q$=$(\partial P/\partial \mu_e)$=$m_s^2\mu_q/(2\pi^2)$. Noting that $x_e\propto\mu_e^3$, we obtain 

\begin{equation}
\left(\frac{\partial P}{\partial x_e}\right)=\frac{m_s^2\mu_q\mu_e}{6\pi^2x_e}   
\end{equation}

Substituting this result in eq.(\ref{csce2}) and using eq.(\ref{BV}), 

\begin{equation}
\label{Nstrange}
N\approx\frac{\sqrt{3}}{2\pi}\frac{g}{c_e}\left(\frac{m_s^2}{\sqrt{\rho-B}}\right) 
\end{equation}

The difference between the adiabatic and equilibrium sound speeds is, in this case, of order $m_s^4/B$. For the case of a strange star, near the surface, $\rho\approx 4B$, so that $N\simeq (g/(2\pi c_e))(m_s^2/\sqrt{B})$, which is in good agreement with the estimate in~\cite{Olinto96}. For neutron stars with a quark core, we may conclude from eq.(\ref{Nstrange}) that the resulting $g$-mode frequency (for low-$l$ values) $\omega_{\rm BV}\sim 100$ Hz, which is very similar to the estimate for non-interacting $(n,p,e^-)$ matter. However, numerical results for realistic models of nuclear and quark matter with interactions show that their \gm\, frequencies are quite different (see Fig.\ref{fig:gmode}). 

\vskip 0.2cm

We now consider the case of a 2-flavor model, namely, the vBag model, as an example of how interactions can induce \gm\,oscillations. The vBag model is a hybrid approach that accounts for scalar interactions and hence chiral symmetry breaking by assuming  bare  quark  masses  and  flavor  dependent  chiral bag constants ($B_{\chi,f}$) to reproduce the proper critical chemical potential  for  each  flavor's  chiral symmetry restoration. Vector interactions are taken into account non-perturbatively in analogy to the  NJL  model. The quark pressure and energy density are given by 

\begin{equation}
P_q=\sum\limits_{f} P_f + B_{dc}\,; \quad \epsilon_q=\sum\limits_{f} \epsilon_f - B_{dc}\,.
\end{equation}

where $B_{dc}$ is the confinement Bag constant, introduced to ensure that quarks are confined in the chirally restored phase. We may take it to be the same for both light flavors. The individual flavor pressure and energy density appearing in the equations above are

\begin{eqnarray}
P_f(\mu_f)=P_{FG,f}(\mu_f^*)+\frac{K_v}{2}n_{FG,f}^2(\mu_f^*)-B_{\chi,f}\\
\epsilon_f(\mu_f)=\epsilon_{FG,f}(\mu_f^*)+\frac{K_v}{2}n_{FG,f}^2(\mu_f^*)+B_{\chi,f}
\end{eqnarray}

with the subscript $FG$ denoting the Free Fermi gas expression. We choose $B_{\chi,u}=B_{\chi,d}$ to avoid sequential restoration. These equations contain the vector repulsion term $\propto K_v$, which comes from vector current-current interactions and is connected to the gluon mass scale in Dyson-Schwinger studies of non-perturbative QCD. The repulsion term is essential to stiffen the quark equation of state, and support compact stars at least as heavy as 2$M_{\odot}$. We will see that it also controls the electron fraction in quark matter, thereby influencing the \gm . The introduction of the vector term also modifies the number densities and chemical potentials as: 

\begin{eqnarray}
\label{eqstar}
\mu_f&=&\mu_f^*+K_v n_{FG,f}(\mu_f^*) \\ 
n_f(\mu_f)&=&n_{FG,f}(\mu_f^*)
\end{eqnarray}

The vBag equation of state can be expressed as:

\begin{equation}
P_q=\frac{1}{3}(\epsilon_q-4\sum\limits_{f}B_{\chi,f})+\frac{4}{3}B_{dc}+\frac{K_v}{3}\,\sum\limits_{f}n_f^2(\mu_f)\,.
\end{equation}

which has a non-barotropic form since $n_f(\mu_f)$ encodes composition information. Charge neutrality requires $(2/3)x_u-(1/3)x_d-x_e$=0, where $x_i$=$n_i/n$ are the quark to baryon number fractions of species $i$. We also impose $\beta$-equilibrium : $\mu_d-\mu_u$=$\mu_e$. Since $x_u+x_d$=3, we can obtain $x_e(n)$ numerically from these conditions. It is useful to note that the depressed cubic  eq.(\ref{eqstar}) has the solution 

\begin{equation}
\overline{{\mu^*}}_f={\rm sinh}\,\left(\frac{1}{3}\,{\rm sinh}^{-1}\,(3\bar{\mu}_f)\right)
\end{equation}

with the scaling $\bar{\mu}$=$(\sqrt{3K_v}/(2\pi))\,\mu$. Subsequently, we obtain an analytic approximation for the electron fraction, which is within 5\% of the numerical result. 

\begin{eqnarray}
\label{xe-tk}
x_e(n)&=&\frac{[\tilde{n}+(2^{1/3}-1)(\pi^2\tilde{n})^{1/3}]^{\,3}}{3\pi^2\tilde{n}}\\
\tilde{n}&=&K_v^{3/2}n
\label{scaling}
\end{eqnarray}

For values in the typical parameter range $K_v\sim$ (2-6) GeV$^{-2}$, the dimensionless baryon density is $\bar{n}\sim$ (0.002-0.02) for densities of relevance to quark matter in compact stars ($n\sim$ (3-6)$n_{\rm sat}$), implying that $x_e\sim$ (0.006-0.014). To obtain the difference in sound speeds from eq.(\ref{csce2}), we note that 
\begin{equation}
\frac{\partial P}{\partial 
x_e}=n^2\frac{\partial }{\partial n}\frac{\partial E}{\partial x_e}=n^2\frac{\partial }{\partial n}\left[\mu_e+\mu_u-\mu_d\right]
\end{equation}
Expressing the chemical potentials $\mu_i$ in terms of the partial fractions $x_i$, we arrive at 
\begin{equation}
\label{d1-tk}
\frac{\partial P}{\partial 
x_e}=-\frac{2}{3}n^2K_v(1-2x_e)
\end{equation}

From eq.(\ref{xe-tk}) and (\ref{scaling}), it follows that 
\begin{equation}
\label{d2-tk}
n\frac{dx_e}{dn}=\frac{2K_v}{(3\pi^2)^{1/3}}(n x_e)^{2/3}
\end{equation}

From eqs.(\ref{d1-tk}) and (\ref{d2-tk}), and to leading order in $x_e$, we find
\begin{equation}
\label{csce-tk}
c_s^2-c_e^2\approx \frac{x_e^{2/3}}{(3\pi^2)^{1/3}}\left(\frac{K_v^2\,n^{8/3}}{\sum\limits_{f}(\epsilon_{FG,f}(\mu_f^*)+\frac{3}{4}K_v\,n_{FG,f}^2(\mu_f^*))}\right)
\end{equation}

To the same order in $x_e$, the \bv\, frequency in this model is 
\begin{eqnarray}
N&\approx&\frac{g\,x_e^{1/3}}{c_e}\times \\ \nonumber
&&\left(\frac{K_v\,n^{4/3}}{\sqrt{(P_q+\epsilon_q)+2\sum\limits_{f}(K_v n_{FG,f}^2(\mu_f^*)+B_{\chi,f})-B_{dc}}}\right)
\end{eqnarray}

Finally, we consider the case of a mixed phase of non-interacting 2-flavor quark matter with nuclear matter, which also implies the existence of \gm\, oscillations, at least in part of the mixed phase. This foreshadows the more realistic numerical treatment of the mixed phase in the next section.

Consider an admixture of two (uncharged) components such as water and air. Even at very low bubble fraction $\chi$, the effective incompressibility is reduced sharply since the density of the mixture is hardly changed, whereas air bubbles significantly reduce the pressure compared to a pure liquid. The resulting equilibrium sound speed is given by~\cite{Wood}

\begin{equation}
\label{cemix}
\frac{1}{c_{\rm mixed}^2}=\frac{(1-\chi)^2}{c_l^2} +\frac{\chi^2}{c_g^2} +\chi(1-\chi)\left(\frac{\rho_g}{\rho_l\,c_l^2}+\frac{\rho_l}{\rho_g\,c_g^2}\right)    
\end{equation}

where $l$ and $g$ stand for the liquid (dense) and gas (void) phase respectively. This assumes that the bubble can exchange heat with the surrounding fluid fast enough during the perturbation to maintain equilibrium. Since the material density $\rho_g\ll \rho_l$, a distinct drop in the equilibrium sound speed is seen at the onset of the mixed phase.   

For a quark-hadron mixed phase, eq.(\ref{cemix}) is not directly applicable since each phase also carries charge such that the system is globally neutral, with the void fraction given by $\chi_q=n_Q^h/(n_Q^h-n_Q^q)$ where $n_Q$ denotes the charge density, and $q$ and $h$ refer to quark and hadron phases. Furthermore, the energy density $\rho_q$ and $\rho_h$ are similar in magnitude unlike for air and water, which would imply $c_{\rm mix}\approx c_h$ for small $\chi$. Taking the two conserved quantities (baryon number and charge) into account, we have

\begin{equation}
\label{ceglen}
c_{\rm mix}^2=\frac{dP_{\rm mix}(\mu_B,\mu_Q)}{d\rho}=\frac{\partial P_{\rm mix}}{\partial\mu_B}\left(\frac{d\mu_B}{d\rho}\right)+\frac{\partial P_{\rm mix}}{\partial\mu_Q}\left(\frac{d\mu_Q}{d\rho}\right)
\end{equation}

Accordingly, an additional contribution to eq.(\ref{cemix}) exists, with the energy densities $\rho_q$ and $\rho_h$ replaced by the respective charge densities $\rho_q^Q$ and $\rho_h^Q$ respectively. As shown in~\cite{Glen} with an explicit quark-hadron mixed phase construction, the first term in eq.(\ref{ceglen}) is continuous with density at the onset of the mixed phase, but the second term involving the charge chemical potential is not. Therefore, we expect, as confirmed by our numerical results presented in sec IV, that $c_{\rm mix}$ has a negative discontinuity at the onset of the mixed phase, in effect lowering the equilibrium sound speed. The physical meaning is that the system is more compressible in a charge separated state, as internal forces have the freedom to rearrange charges between the two phases to minimize the free energy. In effect, global charge neutrality makes the system more compressible. 

Our numerical calculations show a similar sharp drop in the equilibrium sound speed at the onset of the mixed phase, with only a small change in the adiabatic sound speed. Therefore, $c_s^2-c_e^2>0$ and stable $g$-modes can be found. The modes do not survive the entire extent of the mixed phase ($0<\chi<1$). From eq.(\ref{cemix}), $c_{\rm mix}$ increases with $\chi$ then decreases slowly, if we employ the interpretation of $\rho_i$ as charge density and enforce global charge neutrality. On the other hand, the adiabatic sound speed, $c_s$, increases gradually over the mixed phase in the regime $0<\chi<1$. These trends in the two sound speeds lead to $N^2<0$ for some $\chi>\chi_0$, which turns off the restoring force for \gm\, oscillations deep in the core of a hybrid star. These features are corroborated in our numerical results for the mixed phase.

\section{IV. Numerical results for $g$-modes in the mixed phase}

We now examine the \gm\, in the mixed phase using a realistic (but still parameterized) model for the nuclear phase (SL23 EoS) and the vBag model for the 2-flavor quark phase. Details about these EoS are given in~\cite{Wei1}, also   in~\cite{Klahn1,Klahn2,Klahn3,Wei1}. Here, we employ parameters for the vBag model that yield a maximum mass of at least 2$M_{\odot}$, and result in the appearance of a mixed phase of nuclear and two-flavor quark matter in the interior of neutron stars. As explained in our paper~\cite{Wei1} on the masquerade problem, it is possible to consider two phase transitions, the first involving only $(u,d)$ quark matter, and the second involving $s$-quarks at a higher density. One could also choose vBag parameters to have only three-flavor matter in a mixed phase. In either of these two more involved cases, we expect that our conclusions about the rise in \gm\, frequency would not change qualitatively, but we do not study them numerically here.

\begin{figure}[h!]
\centering
\includegraphics[height=65mm]{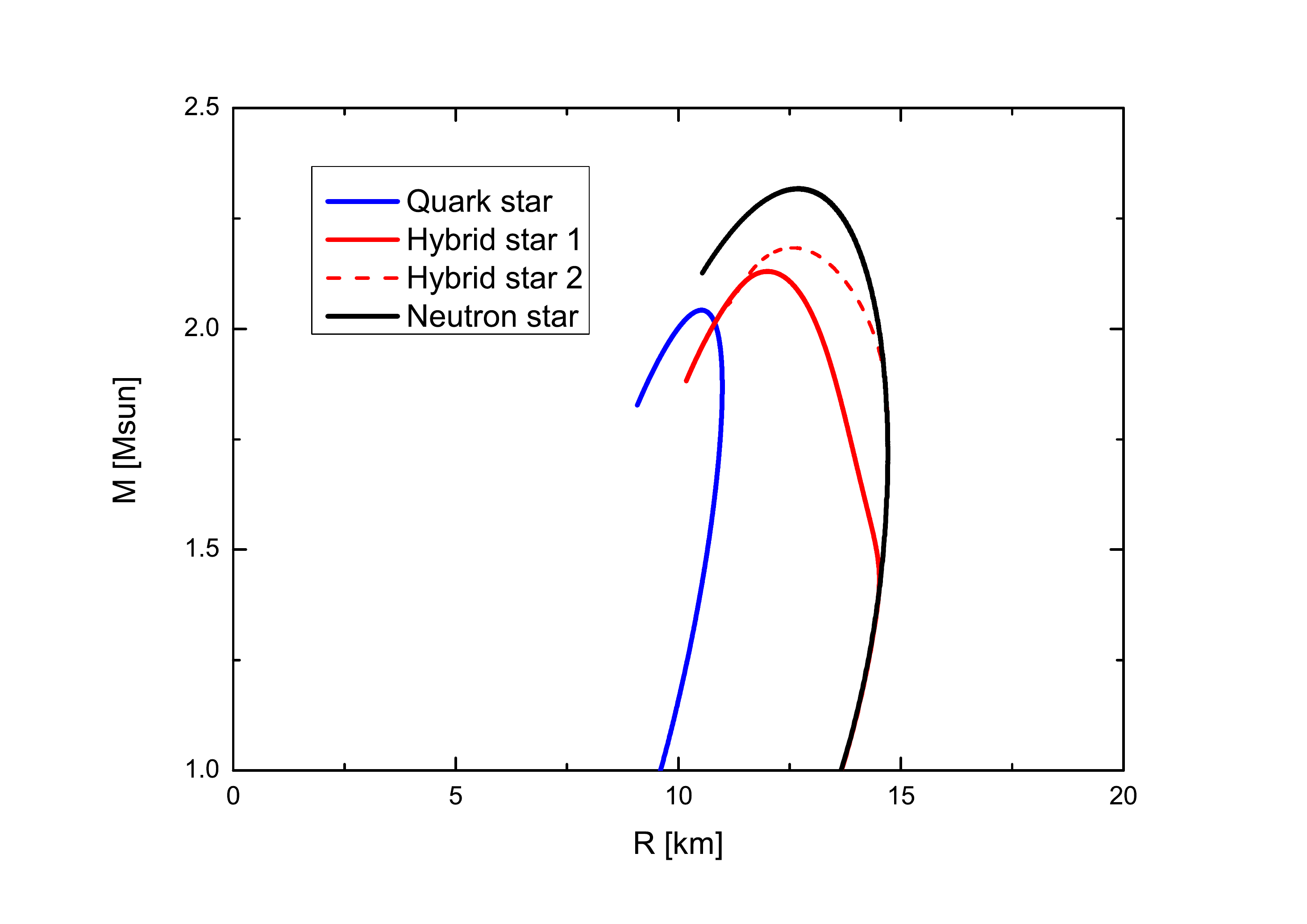}\\
\caption{The mass-radius curve for three different compact star configurations - neutron stars, hybrid stars and quark stars. The parameters of quark matter for the quark/hybrid star 1 are $B_{\rm eff}$=70 MeV fm$^{-3}$ and for hybrid star 2, $B_{\rm eff}$=80 MeV fm$^{-3}$. $K_v$=$6\times 10^{-6}$\,MeV$^{-2}$ in both cases.}
\label{fig:MR}
\end{figure}

\begin{figure}[]
\centering
\includegraphics[height=65mm]{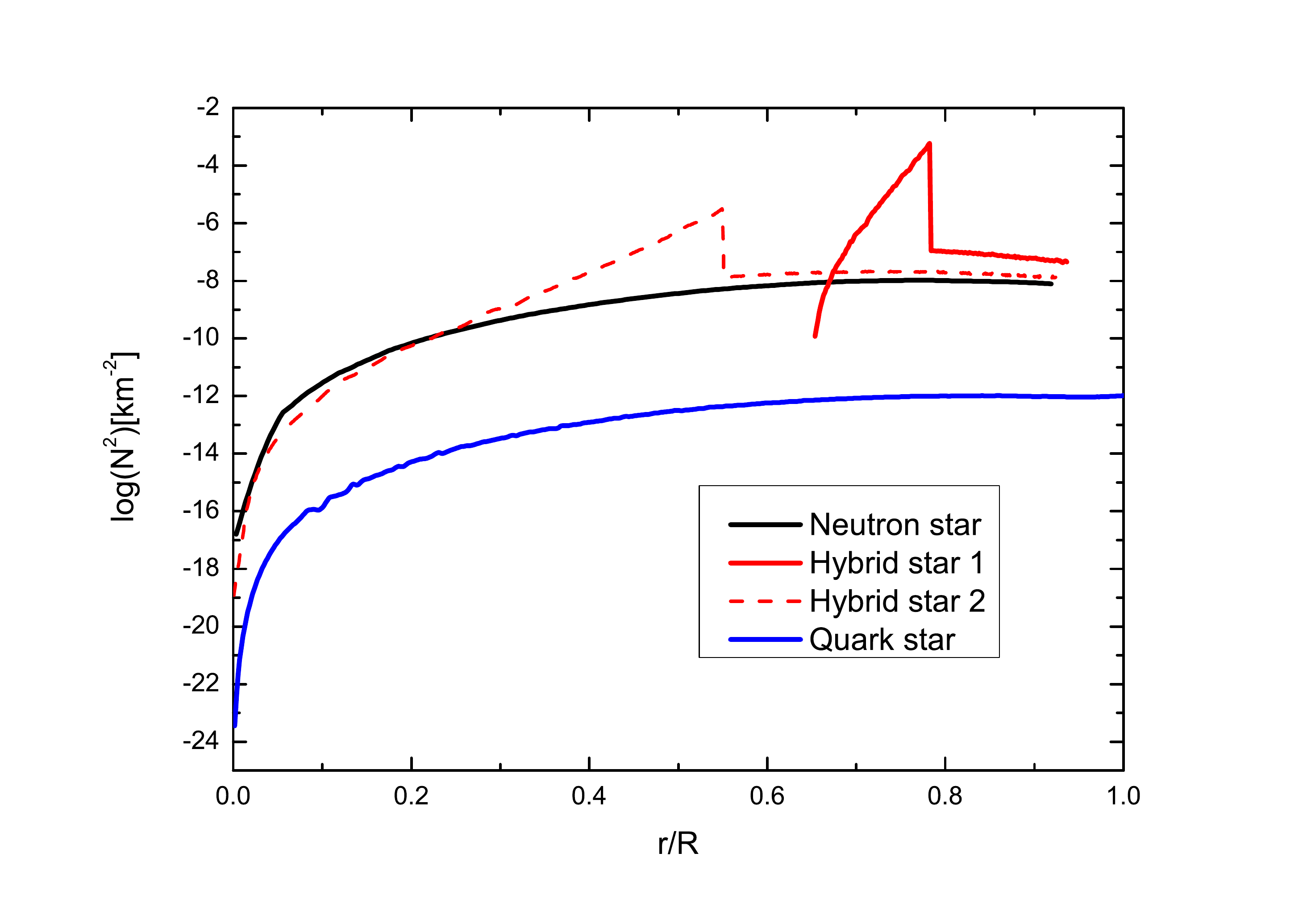}\\
\caption{The local \bv\, frequency for neutron/quark/hybrid stars plotted as a function of the relative distance from the center $r/R$. The parameters are same as in Fig.\ref{fig:MR}. The mass of the neutron star and hybrid stars are $2.1M\odot$ and the mass of the quark star is $1.9M\odot$.}
\label{fig:N2}
\end{figure}

In Fig.\ref{fig:MR}, which shows the mass-radius curve for neutron, quark and hybrid stars with our chosen EoS, we observe that the onset of the softening due to the appearance of the mixed phase happens at higher stellar mass as $B_{\rm eff}$ is increased, contributing to the masquerade effect that was described extensively in~\cite{Wei1}. The vector interaction provides the necessary stiffness to generate masses above 2$M_{\odot}$. Clearly, vBag model parameters can be chosen so as to mask the effect of the phase transition in the mass-radius curve, so we look to the \gm\, signature instead. Fig. \ref{fig:N2} shows how the \bv\, frequency rises abruptly inside the star due to the drastic reduction in the equilibrium sound speed, revealing the onset of the mixed phase. Although this is a continuous phase transition with no sharp density jump, the \gm\, jumps sharply at this point for the reason explained at the end of the previous section. For the hybrid star configuration with $B_{\rm eff}$=70 MeV fm$^{-3}$, the curve terminates at $r/R\approx 0.65$, since $N^2$ turns negative and \gm s are no longer supported.

\begin{figure}[H]
\centering
\includegraphics[height=70mm]{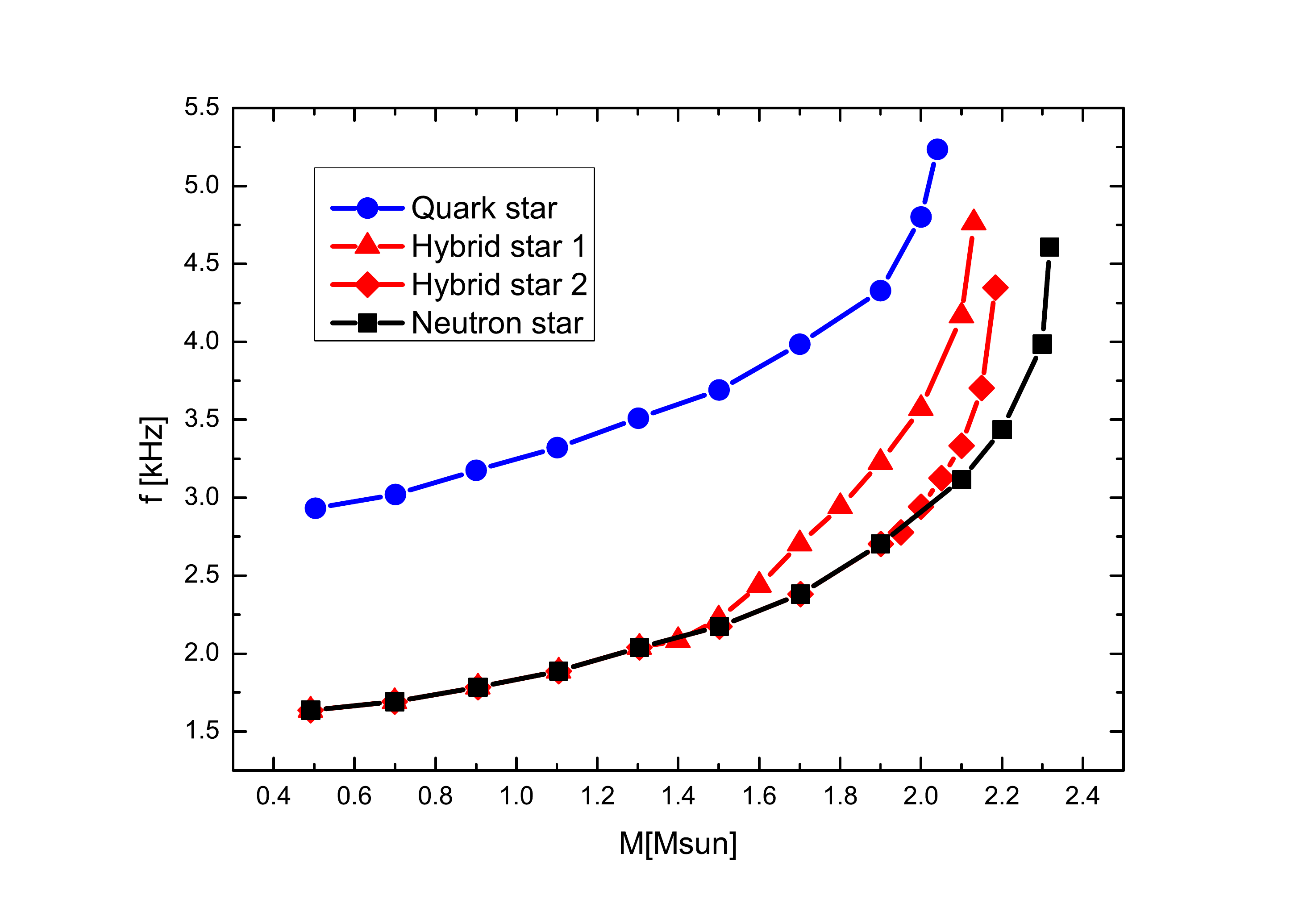}\\
\caption{The Newtonian eigenfrequencies of $f$-modes for the neutron star, hybrid star and quark star as a function of stellar mass. Parameters are same as in Fig.\ref{fig:MR}.}
\label{fig:fmode}
\end{figure}

\begin{figure}[H]
\centering

\includegraphics[height=70mm]{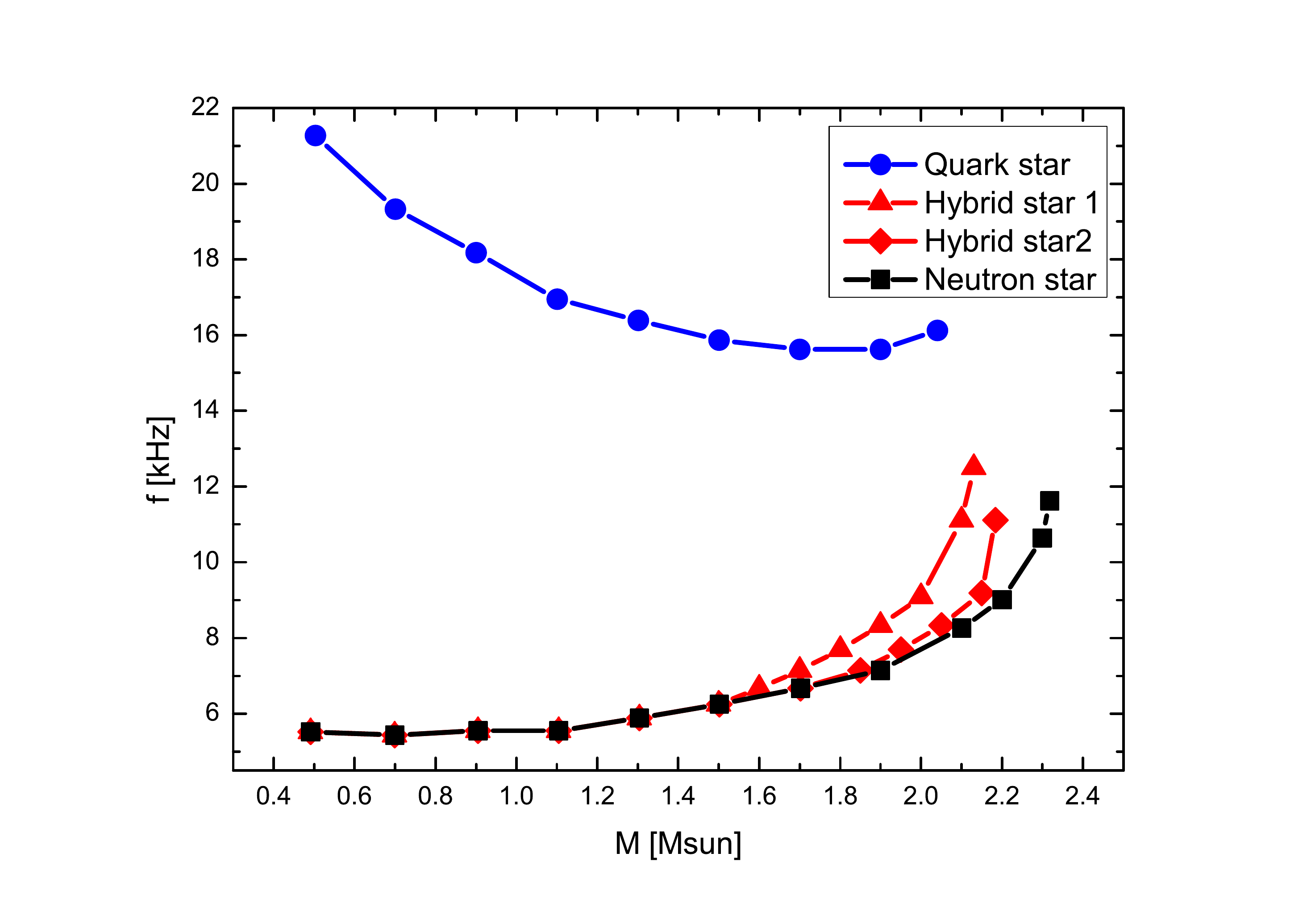}\\
\caption{The Newtonian eigenfrequencies of $p$-modes for the neutron star, hybrid star and quark star as a function of stellar mass. Parameters are same as Fig.\ref{fig:MR}. The trend of the $p$-mode frequency for the quark star, which is opposite that of the neutron and hybrid star, arises because the quark star is self-bound.}
\label{fig:pmode}
\end{figure}

\begin{figure}[H]
\centering
\includegraphics[height=70mm]{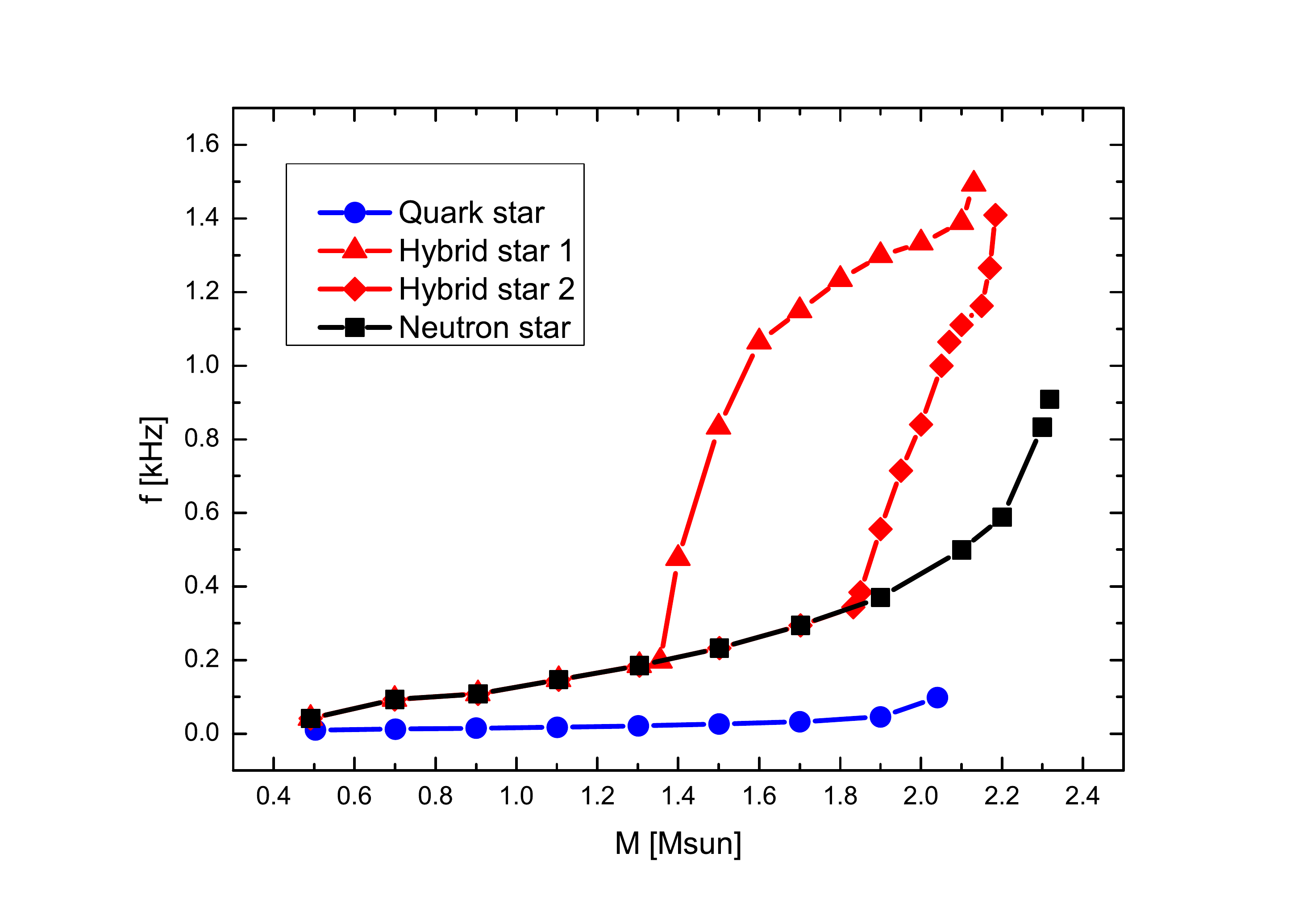}\\
\caption{The Newtonian eigenfrequencies of $g$-modes for the neutron star, hybrid star and quark star as a function of stellar mass. Parameters are same as Fig.\ref{fig:MR}. Note the steep rise in the \gm\, frequency as soon as the mixed phase is favored.}
\label{fig:gmode}
\end{figure}

In Figs. \ref{fig:fmode}, \ref{fig:pmode} and \ref{fig:gmode}, we observe the impact of the phase transition on the fundamental $f,p,g$-modes. 

While the $f$ and \gm s both have frequencies within the sensitivity range of Advanced LIGO/Advanced VIRGO, only the $g$-modes show a trend for hybrid stars that is very different from neutron stars/bare quark stars for any mass above 1.4 $M_{\odot}$. The $f$-modes for hybrid stars appear to interpolate between the neutron star and quark star as we go to increasing mass. The $g$-modes for the hybrid star on the other hand have frequencies much higher than either the neutron star/quark star. Even if the compact star's mass is not measured, a \gm\, frequency of more than about 1 kHz is likely to be supported only in a hybrid star. Within our chosen model, this also constrains the stellar mass to be above 1.6$M_{\odot}$. We also note that lower \gm\, frequencies (0.4-0.8 kHz) could originate from low/intermediate mass hybrid stars or high-mass neutron stars. These two possibilities can be distinguished if the $f$-mode frequency, which is very different for low and high mass stars irrespective of model parameters, is also measured. Therefore, even in the absence of a mass measurement, it is possible to extract information on the interior composition of the compact star such as whether it can support a phase transition to quark matter, using its oscillation spectrum. If the mass is known to better than a few \% and the frequency to better than few tens of Hz (both reasonable limits of precision) we can begin to constrain the parameters of the vBag model which derive from the non-perturbative sector of the strong force. It is also worth noting that while the \gm\, and the $p$-mode frequencies are both quite distinct between neutron stars and pure quark stars, only the \gm\, frequencies would be in the sensitivity band of currently operational detectors. In the next section, we estimate the \gm\, damping time as this affects the likelihood of practically detecting these modes with gravitational wave interferometers.


\section{V. Damping times for the $g$-mode in quark matter}
 In this section, we provide estimates for the damping time of the \gm\, in two-flavor quark matter, to provide some comparison with ordinary neutron stars. These are only order of magnitude estimates that can be refined by utilizing the \gm\, wave functions obtained from the solution of the fluid perturbation equations along the lines of~\cite{Lai94}. However, that is beyond the scope of the current paper. Three sources of damping are identified in~\cite{RG}: neutrino damping (bulk viscosity), damping by shear viscosity and gravitational wave damping (the latter being negative corresponds to mode growth). We address these in turn. Neutrino damping of the \gm\, involves the relaxation of the departure from chemical equilibrium $\delta\mu(n,x_e)$=$\mu_d-\mu_u-\mu_e$ due to the non-equilibrium $\beta$-decay rate. Working at fixed baryon density, we define the typical relaxation timescale through

\begin{equation}
\tau_{\beta}=1/\Gamma_{\rm rel}\,;\quad\dot{\xi}=-\Gamma_{\rm rel}\xi\,;\quad \xi=\frac{\delta\mu}{T}
\end{equation}

Using the relations
\begin{eqnarray}
&&\frac{\partial\delta\mu}{\partial t}=\left(\frac{\partial\delta\mu}{\partial x_e}\right)\dot{x_e}=-\zeta\,\Gamma_{\rm net} \\ 
&& \zeta=-\frac{1}{n}\frac{\partial\delta\mu}{\partial x_e}=-\frac{1}{n}\frac{\partial^2 E}{\partial x_e^2}\,\\
&&\Gamma_{\rm net}=\Gamma_{d\rightarrow u+e+\bar{\nu}_e}-\Gamma_{ u+e\rightarrow d+\nu_e}
\end{eqnarray}
and the expression for $\Gamma_{\rm net}$ from~\cite{Anand}, we obtain

\begin{equation}
\label{betatime}
\tau_{\beta}({\rm yr})\approx 8.2\,T_9^{-4}\left(\frac{n_{\rm sat}}{n}\right)^{2/3}\frac{1}{\left(\delta\mu/{\rm MeV}\right)}
\end{equation}

where $n_{\rm sat}$=0.154 fm$^{-3}$ is the nuclear saturation density, and we have assumed $K_v=4$ GeV$^{-2}$. The magnitude of $\delta\mu$ depends on the amplitude of the oscillation, which is uncertain, but we may assume an upper limit of $\delta\mu\approx 1$ MeV, which corresponds to fluctuations in the chemical potentials at the 1\% level. Since the oscillation timescale for the \gm\, in quark matter is $\sim 0.01-0.1$ seconds, it is clear from eq.(\ref{betatime}) that unless $T>10^{11}$K, the \gm\, is not damped by this mechanism. Even assuming tidal heating during the inspiral, the temperature is insufficient to damp the \gm\, in quark matter through off-equilibrium $\beta$ decays. 
Turning to the damping timescale for shear viscosity, we estimate it to be~\cite{RG,Lai2} 

\begin{equation}
\tau_{\rm visc}({\rm yr})\sim\frac{L^2}{\nu}\approx 1.5\times 10^{3}L_6^2T_9^{5/3}\left(\frac{n_{\rm sat}}{n}\right)^{5/9}\,\, 
\end{equation}

where $\nu$ is the kinematic viscosity, related to the shear viscosity as $\nu$=$\eta/\rho$, and we have used the shear viscosity for quark matter given in~\cite{Heiselberg}, which takes Landau damping into account for the gluons. $L_6$=$L/(10^6\,{\rm cm})$ where $L$ is a typical wavelength scale of oscillation. This timescale is too large to damp the \gm\, by itself unless $T_9 \lesssim 10^{-3}$, i.e, unless $T\lesssim 10^{-6}$, which is the case only for very old neutron stars. 

Finally, we can estimate the effect of the secular instability of the \gm\, in rotating configurations due to gravitational wave emission, also known as the CFS instability~\cite{CFS,FS}. The low frequencies of the \gm\, in quark matter implies that the critical rotation speed at which the CFS instability can be triggered in pure quark stars is $\Omega_s\sim 10-100$ Hz. When the mixed phase enters and the \gm\, frequency rises sharply, stability can be restored. From the analysis in~\cite{Lai2}, we estimate 

\begin{equation}
\label{gw}
\tau_{\rm gw}({\rm yr})\sim\frac{1+{\cal E}}{25}\hat{\omega}_i^{-5}\hat{\omega}_r\frac{R_{10}^4}{M_{1.4}^3}\left(\frac{10^{-4}}{\delta D_{22}}\right)^2\,\,
\end{equation}

where $\hat{\omega}_i$ and $\hat{\omega}_r$ are normalized mode angular frequencies in the inertial and rotating frames respectively, $\delta D_{22}$ is the mass quadrupole and ${\cal E}$ is a sub-leading contribution to the \gm\, energy. Mode instability in the inviscid case sets in when $\omega_i$ turns negative, which happens at a critical spin frequency of $\nu_s \approx 0.68\,\nu_0$~\cite{Lai2}, where $\nu_0$ is the mode frequency. 

Applying eq.~(\ref{gw}) to our quark model EoS for a 1.4$M_{\odot}$ star, for which $\delta D_{22}\approx .0008$ and ${\cal E}\approx 0.7$, we estimate the mode damping timescale to be $\tau_{\rm gw}\sim 10^3$ yrs at zero rotation for a pure quark star and $\tau_{\rm gw}\sim 10^{-2}$ yrs for a hybrid star with a mixed phase quark core. This large difference in damping times is due to the much higher \gm\, frequency in the mixed phase configuration. Taking viscous damping and rotation into account, the overall damping timescale $\tau$, which is given by 

\begin{equation}
\tau = (\tau_{\beta}^{-1}+\tau_{\rm visc}^{-1}+\tau_{\rm gw}^{-1})^{-1}
\end{equation}

implies that the \gm\, can be unstable to gravitational wave emission (i.e, $\tau <0$) in the temperature range $10^8$K$<T<10^9$K for a stellar rotation frequency of about twice the frequency of the \gm\, frequency at zero rotation $\approx 200$ Hz . With decreasing rotation speed, the instability window narrows and ultimately closes. However, at slower rotation speeds, additional sources of damping  such as mutual friction could become important if the quarks are in a superfluid phase.

\section{VI. Conclusions}
Based on our study of core \gm s in compact stars with and without quark matter, we conclude that the frequency of these modes is very sensitive to the presence of a mixed phase containing quarks and hadrons. The equilibrium sound speed drops sharply at the boundary of the mixed phase, raising the local \bv\, frequency and the fundamental \gm\, frequency of the star. If this mode can be resonantly excited during the late stages of binary inspiral, the resulting energy transfer from the orbital motion to the star via tidal coupling can affect the phase of the gravitational waveform, and possibly give a signature of the quark-hadron phase transition in the star. Previous works have examined the accumulated phase error from tidal coupling to the \gm\, for ordinary neutron stars with composition gradients (but no phase transition) and concluded that it is about two orders of magnitude too small to be detected by current detectors~\cite{Lai94,Yu17,Xu17}. However, if one or both stars support a mixed phase of quark-hadron matter, there are really two fluid components inside each star that can be tidally forced. This, and the fact that the spectrum of \gm\, is shifted to higher frequencies and is about 5 times more dense~\footnote{We find 2 distinct \gm s between 100 Hz - 1.5 kHz for neutron stars and 10 distinct modes in the same frequency range for hybrid stars.} than for ordinary neutron stars, imply that more modes can become resonant as the signal sweeps through the bandwidth of the detector and accumulate a larger phase error~\cite{Yu17}. In any case, a final verdict awaits the calculation of the tidal coupling coefficient, energy transfer and the phase error in the case of a hybrid star. 

It is also possible to relate the frequency of the \gm\, to the fraction of quark matter inside the star. 

\vskip 0.5cm
\noindent Table 1: Trends in the \gm\, frequency and volume fraction of the quark phase as a function of stellar and EoS parameters. \vskip 0.2cm
\begin{tabular}{ |C{3cm}|C{1cm}|C{1.85cm}|C{1.75cm}|  }
 \hline
 {\bf EOS parameters}:
 ($B_{\rm eff}$, $K_v$) & {\bf Max. mass} ($M_{\odot}$) & {\bf Volume fraction of quarks} & \gm\, {\bf frequency} (kHz)\\
 \hline
 $B_{\rm eff}$=70\,MeV/fm$^3$ $K_v$=2 GeV$^{-2}$   &  1.93   &67.0\%&   0.5264\\
 \hline
 $B_{\rm eff}$=80\, MeV/fm$^3$ $K_v$=2 GeV$^{-2}$&  1.90  & 45.8\%   &0.8333\\
  \hline
 $B_{\rm eff}$=70\,MeV/fm$^3$ $K_v$=4 GeV$^{-2}$ &2.03 & 59.9\%&  1.1236\\
  \hline
$B_{\rm eff}$=80\,MeV/fm$^3$ $K_v$=4  GeV$^{-2}$   &2.05 & 39.5\%&  0.8333\\
 \hline
$B_{\rm eff}$=70\, MeV/fm$^3$ $K_v$=6 GeV$^{-2}$&   2.13  & 53.1\%&1.4925\\
 \hline
$B_{\rm eff}$=80\,MeV/fm$^3$ $K_v$=6 GeV$^{-2}$& 2.18  & 31.0\%   &1.4085\\
 \hline
\end{tabular}
\label{table1}
\vskip 0.5cm

For the equations of state and parameter pairs ($B_{\rm eff},K_v$) chosen in this paper, the table above shows the volume fraction of quark matter in a hybrid star of maximum mass. For a smaller maximum mass, the \gm\, frequency is higher. For maximum mass above 2$M_{\odot}$, a higher \gm\, frequency implies a larger fraction of quark matter. Given these trends,  the \gm\, frequency can be an indicator of the fraction of quark matter in the star. Note that the \gm\, frequency is identical for maximum mass configurations in the case of $K_v=2$ and $K_v=4$ for the same $B_{\rm eff}$. This is suggestive of a universality in the scaling of the \gm\, for hybrid stars that is intriguing and invites further study.

There are a few physical effects that can alter our results quantitatively, which have not been taken into account. Rotation and full general relativity have not been incorporated at the level of the perturbative analysis. Non-linear mixing between $p$ and \gm s is possible~\cite{Weinberg:2015pxa}, although this effect seems to be disfavored by the data on GW170817~\cite{Reyes:2018bee}. Other nuclear EoS parameterizations and the possibility of strange quarks appearing together or at a higher density than the light quarks would change the sound speed profile and hence the \gm\, frequency. We cannot as yet make any statements on the quantitative impact of the effect proposed here on the gravitational wave signal from binary mergers. However, given the subtle nature of the masquerade problem and our optimism for increased statistics on binary mergers from the next observing run of Advanced LIGO/VIRGO detectors, the \gm\, is a promising diagnostic for the quark-hadron phase transition deserving of further investigation. 

\emph{Acknowledgments.}---W.W. is supported by the Natural Science Foundation
of China under Grant No.11547021 and China Scholarship Council. P.J. is supported by the U.S. NSF Grant No. PHY 1608959. We thank Tobias Fischer, Marc Salinas and Bryen Irving for helpful input and discussions.

\bibliographystyle{unsrt}
\bibliography{paper2}

\end{document}